\documentclass[aps,pra,superscriptaddress,groupedaddress]{revtex4}  % custom
\usepackage{graphicx}  % needed for figures
\usepackage{dcolumn}   % needed for some tables
\usepackage{bm}        % for math
\usepackage{amssymb}   % for math
\usepackage{color}
\usepackage{braket}
\usepackage{amsmath}

\newcommand{\ketbra}[2]{| #1 \rangle \langle #2 |}
\newcommand{\op}[2]{|#1\rangle \langle #2|}
\DeclareMathOperator{\Tr}{Tr}
\DeclareMathOperator{\tr}{Tr}

\newcommand{\mbb}{\mathbb}
\newcommand{\mc}{\mathcal}

\newcommand{\mf}[1]{\mathfrak{ #1}}

\newtheorem{lemma}{Lemma}
\newtheorem{theorem}{Theorem}
\newtheorem{prop}{Proposition}
\newtheorem{definition}{Definition}

\begin{document}

% the following line is for submission, including submission to the arXiv!!
%\hspace{5.2in} \mbox{Fermilab-Pub-04/xxx-E}

\title{One-shot assisted concentration of coherence}
\author{Madhav Krishnan V}
\affiliation{Centre for Quantum Software and Information, University of Technology Sydney, NSW 2007, Australia}
\author{Eric Chitambar}
\affiliation{Department of Physics and Astronomy, Southern Illinois University, Carbondale, Illinois 62901, USA}
\author{Min-Hsiu Hsieh}
\affiliation{Centre for Quantum Software and Information, University of Technology Sydney, NSW 2007, Australia}

\date{\today}

\begin{abstract}
We find one-shot bounds for concentration of maximally coherent states in the so called assisted scenario.  In this setting, Bob is restricted to performing incoherent operations on his quantum system, however he is assisted by Alice, who holds a purification of Bob's state and can send classical data to him.  We further show that in the asymptotic limit our one-shot bounds recover the previously computed rate of asymptotic assisted concentration. 
\end{abstract}

%\pacs{}
\maketitle

\section{Introduction}
Resource theories have been a powerful tool in developing our understanding of Quantum Information Theory (QIT). For instance QIT can be seen as theories of interconversion of different resources \cite{devetak_2008_resource}. Reource theories have also allowed us to quantify the role of various quantum features, such as entanglement, in the performance of quantum computational tasks \cite{horodecki_2009_quantum, plenio_2007_introduction}. Outside of information theory, the resource-theoretic approach has found application in the study of quantum thermodynamics \cite{gour_2015_resource, brandao_2010_generalization} and shared reference frames \cite{gour_2008_resource}, among many others. General measures such as the \textit{relative entropy of resource} can be applied in different resource theories and carry analogous operational interpretations in each \cite{Horodecki-2002a, Horodecki-2013a, Brandao-2015a, anshu_2017_quantifying}.  

The focus of this paper is the resource theory of quantum coherence.  The fact that coherent superpositions of quantum states are valid physical states is an essential feature of quantum mechanics, and it provides advantages of quantum computation over it's classical counterpart. The phenomenon of coherent superposition states has received renewed focus through the lens of a quantum resource theory  \cite{baumgratz_2014_quantifying, aberg_2006_quantifying, levi_2014_quantitative, chitambar_2016_comparison, du_2015_conditions, winter_2016_operational, yadin_2016_quantum}. See also \cite{streltsov_2017} for a detailed review. In the resource theory of coherence, a state is considered resourceful if it is non-diagonal in a particular fixed basis. In other words, all free states have the form $\delta = \sum\limits_i \delta_i \op{i}{i} $, where $\{\ket{i}\}_i$ is some fixed basis known as the \textit{incoherent basis}. Several different families of free operations have been proposed in the literature, and they all share in the property of mapping the set of free states onto itself. Quantification of the amount of coherence in a quantum state is achieved through several coherence monotones, one such coherence monotone being the relative entropy of coherence $C_r(\cdot)$. The conditions which allow transformation between different states via free operations is an important operational question that has been answered for pure states but is still unknown for mixed states \cite{winter_2016_operational}. There also exists a maximally resourceful state and it's optimal interconversion with arbitrary states defines the task of coherence concentration, coherence distillation and coherence dilution. Interestingly, when the free operations are identified as the so-called class of \textit{incoherent operations} \cite{baumgratz_2014_quantifying}, the relative entropy of coherence turns out to be the optimal rate for distilling maximally coherent states, thereby providing an operational interpretation of the measure.  The aim of this work is to compute the rate for generating maximally coherent states on one system under the assistance of a second party.

There is a strong similarity between the resource theories of coherence and entanglement, and some of these connections have been pointed out in \cite{chitambar_2016_relating,streltsov_2015_measuring,streltsov_2016_entanglement,zhu_2017_operational}
. The equivalence in structure between the coherence of assistance and the entanglement of assistance was exploited in \cite{chitambar_2016_assisted} to find the asymptotic coherence of assistance. Inspired by previous work on the problem of one-shot, or single-copy, assisted entanglement concentration \cite{buscemi_2013_general}, in this paper we bound the \textit{one-shot assisted coherence concentration}.  In the assisted concentration scenario, Alice and Bob share a bipartite pure state $\ket{\psi}^{AB}$ and the goal is to maximize the rate of concentration of unit maximally coherent states (MCS) $\ket{\Phi_2} = \frac{1}{\sqrt{2}}\left( \ket{0} + \ket{1} \right) $ on Bob's side,
while Bob is restricted to using incoherent operations and one-way communication is allowed from Alice to Bob. The ideal assisted concentration rate in the asymptotic setting $C_c(\psi^{AB})$, i.e., when Alice and Bob share many copies of the state $\ket{\psi}^{AB}$, is known to be equal to coherence of assistance \cite{chitambar_2016_assisted}. While this rate is achievable with many copies of the state, in realistic scenarios resources are limited.  Thus a more practical question is the following: if we allow for some bounded error in the process, how many copies of a maximally coherent state can we generate from just a single copy of the given pure state $\psi^{AB}$?  While this question has been answered for concentration and dilution in the unassisted setting  \cite{regula_2017_one, zhao_2018_one, Zhao_winter_2018}, it has remained an open question for the one-shot assisted concentration paradigm, and it is one that we answer in this paper.

In section \ref{def_section} we clarify notation and present definitions for the quantities we use in this paper.  In section \ref{pure_state_section} we derive bounds on the one-shot (unassisted) concentration of MCS from an arbitrary pure state.  While this problem has been previously solved in \cite{regula_2017_one}, our approach uses different techniques.  In section \ref{ensemble_section} we generalize these bounds to get the average rate of concentration from an ensemble of pure states and in section \ref{asymp_section} we show that in the asymptotic limit we recover the appropriate rate. Finally we present our conclusions in section \ref{conclusion_section}.

\section{Definitions and Notations}\label{def_section}

We fix a particular basis $\lbrace \ket{i} \rbrace_i $ as the incoherent basis and let $\mathcal{I}$ denotes the set of states  which are incoherent (diagonal) in this basis.
We will use the fidelity measure defined as,
\begin{equation}
F(\rho, \sigma) := \tr \left( \sqrt{\sqrt{\sigma} \rho \sqrt{\sigma} } \right) =  \| \sqrt{\rho}\sqrt{\sigma} \|_1.
\end{equation}
The following lemmas are well-known.

%\begin{lemma}\label{pure_gentle_fid_lemma}
%For any pure state $\ket{\psi}$ and any $\epsilon \geq 0$, if $0 \leq P \leq \mathbb{I}$ is an operator such that $\Tr(P\psi) \geq 1 - \epsilon$, then,
%\begin{equation}
%F(\ket{\psi}, \sqrt{P}\ket{\psi}) \geq 1 - \sqrt{\epsilon} .
%\end{equation}
%\end{lemma}
%Proof: see \cite{buscemi_2013_general}.

%\begin{lemma} \label{gentle_fidelity_lemma}
%For any normalized state $\rho$ and any $\epsilon \geq 0$ and any $0 \leq P \leq \mathbb{I}$, if $\Tr(P\rho) \geq 1 - \epsilon$ then,
%\begin{equation}
%F(w, \rho) \geq 1  - 2\sqrt{\epsilon},
%\end{equation}
%where $w := \frac{\sqrt{P}\rho\sqrt{P}}{\Tr[P\rho]}$.
%\end{lemma}
%Proof : see \cite{buscemi_2013_general}

\begin{lemma} \label{op_diff_lemma}
For any self-adjoint operator A and B and any positive operator $0 \leq P \leq \mathbb{I}$,
\begin{equation}
\Tr(P(A - B)) \leq \Tr(A - B)_+ \leq \|A - B\|_1,
\end{equation}
where $(X)_+$ denotes the positive part of the operator $X$.
\end{lemma}
Proof: see \cite{bowen_2006_beyond}
\begin{lemma}\label{gentle_measurement_lemma}
For any state $\rho$ and an operator $0 \leq \Lambda \leq \mathbb{I}$ such that $\Tr(\Lambda\rho ) \geq 1 - \epsilon$ then,
\begin{equation}
\| \rho - \sqrt{\Lambda }\rho \sqrt{\Lambda} \|_1 \leq 2\sqrt{\epsilon}
\end{equation}
\end{lemma}
Proof: see \cite{ogawa_2002_new}, \cite{winter_1999_coding}.

We also define the following entropic quantities: for any $\rho, \sigma \geq 0$ and any $0 \leq P \leq \mathbb{I}$, and $\alpha \in (0, \infty) \setminus \lbrace 1 \rbrace $  ,
\begin{equation}\label{Eq:S_P_alpha_def_eq}
S^{P}_{\alpha}(\rho \| \sigma) = \frac{1}{\alpha - 1}\log_2 \Tr \left[ \sqrt{P} \rho^{\alpha} \sqrt{P} \sigma^{1 - \alpha}  \right].
\end{equation}
Notice that for $P = \mathbb{I}$, this reduces to the relative R\'enyi entropy. We will be often using the quantity,
\begin{equation}
S_{0}^{P}(\rho \| \sigma) = \lim_{\alpha \rightarrow 0} S_{\alpha}^{P}(\rho | \sigma) = -\log_2 \Tr \left[\sqrt{P} \Pi_{\rho} \sqrt{P} \sigma \right],
\end{equation}
where $\Pi_{\rho}$ is the projector unto the support of $\rho.$ Notice that the quantity,
\begin{equation}
S^{\mathbb{I}}_{0}(\rho \| \sigma) = S_{0}(\rho \| \sigma) = -\log_2(\Tr \Pi_{\rho}\sigma)
\end{equation}
is the relative R\'enyi entropy of order 0. The relative entropy of coherence is defined as,
\begin{equation}
C_r(\rho) := \min_{\delta \in \mc{I}} S(\rho \| \delta) = S(\Delta(\rho)) - S(\rho),
\end{equation}
where $S(\cdot \| \cdot) \equiv S_1^{\mbb{I}}(\cdot \| \cdot) \equiv S_1(\cdot \| \cdot)$ is the relative entropy and $S(\cdot)$ is the von-Neumann entropy.  We use $S_0(\rho \| \sigma )$ to define the min-entropy of coherence as,
\begin{equation}\label{Eq:C_min_def_eq}
C_{min}(\rho) = \min_{\sigma \in \mathcal{I}} S_{0}(\rho \| \sigma).
\end{equation}
where $\mathcal{I}$ is the set of incoherent states.
We also define the min-entropy as,
\begin{equation}
S_{min}(\rho) = -\log_2(\lambda_{max}(\rho)),
\end{equation}
where $\lambda_{max}(\rho)$ is the largest eigenvalue of $\rho$.
To define smoothed versions of these entropic quantities we define the $\epsilon$-close ball for any state $\rho$ and $\epsilon \geq 0$ as,
\begin{equation}
b(\rho, \epsilon) = \lbrace \sigma : \sigma \geq 0, \Tr[\sigma] = 1, F^{2}(\rho, \sigma) \geq 1 - \epsilon \rbrace.
\end{equation}
Similarly we define a $\epsilon$-close ball of pure states
\begin{equation}
b_{*}(\rho, \epsilon) = \lbrace \ket{\psi} : \psi \in b(\rho, \epsilon)\rbrace.
\end{equation}
%We can alternatively define an operator bubble around $\rho$ as,
%\begin{equation}
%\mathfrak{p}(\rho, \epsilon) = \left\lbrace P : 0 \leq P \leq \mathbb{I}, \Tr[P\rho] \geq 1 - \epsilon \right\rbrace.
%\end{equation}
We define the smoothed versions of $S_{min}( \cdot )$  as follows : for any given $\epsilon \geq 0$, the smoothed min-entropy is defined as,
\begin{equation}
S^{\epsilon}_{min}(\rho) = \max_{\overline{\rho} \in b(\rho, \epsilon)} S_{min}(\overline{\rho}).
\end{equation}
We define a pure state smoothed version of this quantity as,
\begin{equation}\label{s_min_*_def_eqn}
S_{min}^{* \epsilon}(\rho) = \max_{\psi \in b_{*}(\rho, \epsilon)} S_{min}(\psi),
\end{equation}
where the smoothing is confined to the pure state ball $b_*(\rho, \epsilon)$.
% We define smoothed versions of $C_{min}$ as
%\begin{equation}
%C_{min}^{\epsilon}(\rho) = \max_{\overline{\rho}  \in \mathfrak{b}(\rho, \epsilon)} \min_{\sigma \in \mathcal{I}} S_0(\overline{\rho} \| \sigma).
%\end{equation}
%We also define the operator smoothed version of $C_{min}$ as
%\begin{equation}\label{c_tilde_def_eqn}
%\widetilde{C}_{min}^{\epsilon}(\rho) = \max_{P  \in \mathfrak{p}(\rho, \epsilon)} \min_{\sigma \in \mathcal{I}} S^P_0(\rho \| \sigma).
%\end{equation}
The asymptotic coherence of assistance for a state $\rho^B$ is defined as \cite{chitambar_2016_assisted}
\begin{equation}\label{asy_coa_def_eq}
 D_a(\rho^B)  :=  \max_{\substack{\lbrace p_i, \psi_i \rbrace_i :\\ \sum\limits_i p_i \psi_i = \rho^B } } \sum\limits_i p_i C_r(\psi_i^{AB}) = S(\Delta(\rho^B)) ,
 \end{equation} 
 where $C_r(\rho)$ is the relative entropy of coherence, $S(\rho)$ is the von-Neumann entropy and $\Delta$ is the dephasing operation in the fixed reference basis. We define the one-shot assisted coherence concentration as,
 
\begin{equation}
 C_{\mathcal{O, \epsilon}}^{A | B}(\ket{\psi}^{AB}) := \max_{\Lambda \in \mathcal{O}}\lbrace \log_2 M | F^2(\Lambda^{AB \rightarrow B^{\prime}}(\ket{\psi}^{AB}), \Phi_M^{B^{\prime}}) \geq  1 - \epsilon\rbrace,
\end{equation}
where $\mathcal{O}$ is the set of local quantum-incoherent operations with one-way classical communication (LQICC-1), $\epsilon \geq 0$, $F(\rho, \sigma)$ is the fidelity and,
\begin{equation}
\ket{\Phi_M} = \frac{1}{\sqrt{M}}\sum\limits_i \ket{i} ,
\end{equation}
is the maximally coherent state of rank $M$. The optimal procedure for Alice to assist Bob would be to perform some POVM $\lbrace P^A_i \rbrace_i$ on her part of the state and communicate the result to Bob who would then apply an incoherent operation $ \Lambda_i $ depending on Alice's outcome. Then the optimal rate must be equal to the following quantity :
\begin{equation}\label{c_a_def_1}
C_{a}(\rho^{B}, \epsilon) := \max_{\lbrace P^A_i \rbrace_i}\max_{M \in \mathbb{N}} \left\lbrace \log_2 M : \max_{\lbrace \Lambda^{B}_i \rbrace_i} F^{2}\left( \sum\limits_{i}p_i \Lambda_i^{B}(\rho_{i}^{B}), \Phi_{M}^{B^{\prime}} \right) \geq 1 - \epsilon  \right\rbrace ;
\end{equation}
which we call the one-shot coherence of assistance, where $p_i \rho_i^B = \Tr_A((P_i^A \otimes \mathbb{I}^B ) \psi^{AB})$.  
The one-shot coherence of assistance can be equivalently defined as,
\begin{equation}\label{c_a_def_2}
C_{a}(\rho^{B}, \epsilon) := \max_{\lbrace p_i, \psi^{B}_i  \rbrace_i}\max_{M \in \mathbb{N}} \left\lbrace \log_2 M : \max_{\lbrace \Lambda^{B}_i \rbrace_i} F^{2}\left( \sum\limits_{i}p_i \Lambda_i^{B}(\psi_{i}^{B}), \Phi_{M}^{B^{\prime}} \right) \geq 1 - \epsilon  \right\rbrace ,
\end{equation}
where $\rho^B = \sum\limits_i p_i \psi_i^B$, since without loss of generality, the maximization over POVMs $\lbrace P_i^A \rbrace_i$ can be restricted to rank-1 POVMs and this is equivalent to preparing any ensemble on Bob's side consistent with his reduced state $\rho^B$ \cite{buscemi_2013_general}. Operationally the concentration task can be split into two parts; Alice prepares an optimal pure state ensemble $\lbrace p_i, \psi_i^B \rbrace_i$ by performing a suitable measurement and communicates the index $i$ to Bob. Bob then performs an optimal incoherent operation on this state to distil the maximally coherent state. Then our task is reduced to finding the optimal rate of distilling the optimal pure state ensemble which will be the best achievable rate on average.

\section{Pure state concentration}\label{pure_state_section}
We will now derive bounds for the one-shot pure state concentration of MCS. The pure one-shot coherence concentration rate for a pure state $\psi$, a set of incoherent operations $\mathcal{O}$ and $\epsilon \geq 0$ is defined as :
\begin{equation}
C_{c}(\psi , \epsilon) := \max_{M \in \mathbb{N}} \left\lbrace \log_2 M : \max_{\Lambda \in \mathcal{O}} F^2(\Lambda(\psi), \Phi_{M}) \geq 1- \epsilon \right\rbrace.
\end{equation}

\vspace{0.2cm}
\noindent\fbox{\begin{minipage}{\dimexpr\textwidth-2\fboxsep-2\fboxrule\relax}
\centering

\begin{theorem}
\label{Thm:pure_state}
For any pure state $\psi$ and $\epsilon \geq 0$
\begin{equation}\label{pure_state_result_eqn}
 S_{min}^{* \epsilon}(\Delta(\psi))  - \delta \leq C_{c}(\psi, \epsilon) \leq S_{min}^{*,2\epsilon}(\Delta(\psi)),
\end{equation}
where $0 \leq \delta \leq 1$ is a number which ensures the lower limit is the logarithm of an integer.
\end{theorem}
\end{minipage}}
\vspace{0.2cm}

Proof: It is known that if $\Delta(\Phi_{M}) \succ \Delta(\psi) $, where the notation $p \succ q$ indicates that $p$ majorizes $q$, then  there exists an incoherent operation $\Lambda$ such that $\Lambda(\psi) = \Phi_{M}$ \cite{winter_2016_operational}. Let $spec(\Delta(\Phi_{M})) = (\frac{1}{M}, \frac{1}{M},...,\frac{1}{M})$ and $spec(\Delta(\psi)) = (\psi_1, \psi_2, ,..., \psi_d)$ 
Then the majorization condition implies that,
\begin{equation}\label{majorization_equation}
\sum\limits_{i = 1}^{k}\frac{1}{M} \geq \sum\limits_{i = 1}^{k} \psi^\downarrow_i, \hspace{1cm}\forall k, d,
\end{equation}
where the $\psi^\downarrow_i$ are the $\psi_i$ in a monotonically decreasing order.  Notice that in this case $\frac{1}{M} \geq \psi_{\max} \equiv \max\limits_j \psi_{j}$ is sufficient to imply the majorization condition in equation \eqref{majorization_equation} and ensuring the existence of an SIO that achieves the desired transformation.  This implies that $\Lambda(\psi)=\Phi_M$ for any $M$ such that $S_{\min}(\Delta(\psi))=-\log\lambda_{\max}\geq \log M$.  In particular $M=\lfloor 2^{S_{min}(\Delta(\psi))}\rfloor$ is always achievable.
%\begin{equation}
%\begin{split}
%&\log_2\left(\frac{1}{\psi_{max}}\right) \geq \log_2M ,\\
%\implies& S_{min}(\Delta(\psi)) \geq \log_2M ,\\
%\implies& 2^{S_{min}(\Delta(\psi))} \geq M ,\\
%\implies& \lfloor 2^{S_{min}(\Delta(\psi))} \rfloor \geq M \hspace{1cm} \because M \in \mathbb{N} ,\\
%\implies& \log_2 \lfloor 2^{S_{min}(\Delta(\psi))} ,\rfloor \geq \log_2 M.
%\end{split}
%\end{equation}
%The maximum rate that can be achieved without error using the above protocol is then $\log_2 M_{max} =  %\log_2 \lfloor 2^{S_{min}(\Delta(\psi))} \rfloor$. This implies that
%\begin{equation}
%C_{c}(\psi, 0) \geq  \log_2 \lfloor 2^{S_{min}(\Delta(\psi))} \rfloor.
%\end{equation}
Consequently, for any pure state $\overline{\psi} \in b_{*}(\psi, \epsilon)$ there exists an SIO operation $\Lambda$ such that $\Lambda(\overline{\psi})=\Phi_{\overline{M}}$ for $\overline{M}=\lfloor 2^{S_{min}(\Delta(\overline{\psi}))}\rfloor$.  Due to the monotonicity of fidelity under positive trace-preserving maps we have,
\begin{equation}
\begin{split}
1 - \epsilon \leq 1 - \epsilon^2 &\leq F^2(\psi, \overline{\psi}), \\ 
&\leq F^2(\Lambda(\psi), \Lambda(\overline{\psi}) ) ,\\
&= F^2(\Lambda(\psi) , \Phi_{\overline{M}} ).
\end{split}
\end{equation}
Hence, $C_{c}(\psi, \epsilon) \geq \log_2\overline{M} $ for any state $\overline{\psi} \in b_{*}(\psi, \epsilon)$, or
\begin{equation}
C_{c}(\psi, \epsilon) \geq \max_{\overline{\psi} \in b_{*}(\psi, \epsilon)}\log_2\lfloor 2^{S_{min}(\Delta(\overline{\psi}))} \rfloor.
\end{equation}
Using the definition of $S_{min}^{* \epsilon}(\rho)$ in equation \eqref{s_min_*_def_eqn} we can write,
\begin{equation}
C_{c}(\psi, \epsilon) \geq \log_2\lfloor 2^{S_{min}^{* \epsilon}(\Delta(\psi))} \rfloor.
\end{equation}

For the converse, let $M$ be the maximum of all $\epsilon$-achievable rates for concentration of the pure state $\psi$, i.e., there exists an incoherent operation $\Lambda$ such that $F^2(\Lambda(\psi), \Phi_M) \geq  1 - \epsilon$. Note that for any incoherent state $\gamma \in \mc{I}$ we
\begin{equation}\label{phi_m_eq}
\Phi_M \Lambda(\gamma) \Phi_M = \frac{1}{M} \Phi_M,
\end{equation}
since $\delta \in \mc{I}$ implies $\Phi_M \delta \Phi_M = \frac{1}{M}\Phi_M$.  Multiplying both sides of equation \eqref{phi_m_eq} with $\Lambda(\psi)$ and taking the trace gives,
\begin{equation}\label{phi_M_eqn_step}
\begin{split}
\tr(\Lambda(\psi) \Phi_M \Lambda(\gamma) \Phi_M ) &= \frac{1}{M}\tr(\Lambda(\psi)\Phi_M), \\
&\leq \frac{1}{M},
\end{split}
\end{equation}
where in the second line we have used the fact that $\Lambda(\psi) \leq \mbb{I}$. Continuing from equation \eqref{phi_M_eqn_step} we have,
\begin{equation}
\begin{split}
\log_2M &\leq -\log_2\tr(\Phi_M \Lambda(\psi) \Phi_M \Lambda(\gamma)  ) ,\\
&= -\log_2\tr(\Lambda^*(\Phi_M \Lambda(\psi) \Phi_M ) \gamma  ),
\end{split}
\end{equation}
where $\Lambda^*$ is the dual map of $\Lambda$ such that $\Tr(X \Lambda(\rho) ) = \tr(\Lambda^*(X) \rho)$. Defining $Q := \Lambda^*(\Phi_M \Lambda(\psi) \Phi_M )$ we have,
\begin{equation}
\begin{split}
\log_2M &\leq -\log_2\tr(Q\gamma) ,\\
&\leq -\log_2\tr(\sqrt{Q} \psi \sqrt{Q} \gamma),\\
&\leq -\log_2\tr(\overline{\psi}\gamma),
\end{split}
\end{equation}
where we use the fact that $\sqrt{Q} \psi \sqrt{Q} \leq Q$ and we have introduced the normalized state $\ket{\overline{\psi}} \equiv\frac{\sqrt{Q} \ket{\psi}}{\sqrt{\bra{\psi} Q\ket{\psi}}}$.  Since $\gamma$ is an arbitrary incoherent state, we thus have
\begin{equation}\label{smin_eqn_step}
\begin{split}
\log_2M &\leq \min_{\gamma \in \mc{I}} \left\lbrace -\log_2\tr( \overline{\psi}\gamma ) \right\rbrace , \\
&=-\log_2(\lambda_{max}(\Delta(\overline{\psi})) ,\\
&= S_{min}(\Delta(\overline{\psi})).
\end{split}
\end{equation}
We will now show that $\overline{\psi} \in b_*(\psi, 2\epsilon)$. Note that

\begin{equation}\label{tr_qpsi_eqn_step}
\begin{split}
\sqrt{\tr(Q\psi)} &= \sqrt{\tr(\Phi_M \Lambda(\psi) \Phi_M \Lambda(\psi)))} = \braket{\Phi_M | \Lambda(\psi) | \Phi_M } ,\\
&= F^2(\Lambda(\psi) , \Phi_M ) \geq 1 - \epsilon.
\end{split}
\end{equation}
and
\begin{equation}
\label{Eq:Fidelity-pure-approx}
\begin{split}
F(\psi, \overline{\psi}) &= \frac{\bra{\psi}\sqrt{Q}\ket{\psi}}{\sqrt{\bra{\psi}Q\ket{\psi}}} , \\
 &\geq \frac{\bra{\psi}Q\ket{\psi}}{\sqrt{\bra{\psi}Q\ket{\psi}}} , \\
 & = \sqrt{\Tr(Q\psi)} ,\\
 & > 1  - \epsilon ,
\end{split}
\end{equation}
where the last inequality follows from equation \eqref{tr_qpsi_eqn_step}. This implies that,
\begin{equation}
F^2(\psi, \overline{\psi}) > 1 - 2\epsilon
\end{equation}
and $\overline{\psi} \in b_*(\psi, 2\epsilon)$. From equation \eqref{smin_eqn_step} we can write,
\begin{equation}
\begin{split}
\log_2M &\leq S_{min}(\Delta(\overline{\psi})) ,\\
&\leq S_{min}^{*, 2\epsilon}(\Delta(\psi)) ,
\end{split}
\end{equation}
thus proving the theorem.

%
%For the other direction, let M be the maximum of all achievable rates of coherence concentration for the pure state $\psi$. Then there exists an incoherent operation $\Lambda$ such that $F^2(\Lambda(\psi), \Phi_{M}) \geq 1 - \epsilon$. Then by definition,
%\begin{equation}
%\begin{split}
%C_{c}(\psi, \epsilon) &= \log_2 M \\
%&= C_{min}(\Phi_{M}) \hspace{3.5cm}  \\
%&\leq C^{\epsilon}_{min}(\Lambda(\psi)) \hspace{3.31cm} \because \text{lemma \ref{C_min_smooth_lemma}} \\
%&\leq \widetilde{C}_{min}^{2\sqrt{\epsilon}}(\psi) \hspace{3.84cm} \because \text{lemma \ref{C_Min_dataprocessing_lemma}} \\
%&\leq S_{min}^{*, \epsilon^{\prime}}(\Delta(\psi)) - \log(1 - 2\sqrt{\epsilon}) \hspace{0.97cm} \because \text{lemma \ref{C_min_S_min_lemma}}
%\end{split}
%\end{equation}
%where $\epsilon^{\prime} = 2^{\frac{5}{4}}\epsilon^{\frac{1}{8}}$. Thus proving the theorem.

We note that our Theorem \ref{Thm:pure_state} is essentially equivalent to the result given in \cite{regula_2017_one}.  Using the theory of distillation norms, the authors of Ref. \cite{regula_2017_one} have shown the one-shot pure state concentration of coherence to be 
\begin{equation}
\label{Eq:Regula-result}
C_c(\psi , \epsilon) = \min_{\sigma \in \mc{I}}D_H^{\epsilon}(\psi \| \sigma ) - \delta ,
\end{equation}
where $D_H^{\epsilon}(\psi \| \sigma)$ is the smoothed hypothesis testing relative entropy.  That is, $D_H^\epsilon(\psi \|\sigma)=-\log \min\{\tr[\sigma M] : 0\leq M\leq\mbb{I},\; F^2(\psi,M)>1-\epsilon\}$.  By applying Sion's minimax theorem \cite{Sion-1958a}, we see that Eq. \eqref{Eq:Regula-result} reduces to
\begin{equation}
\label{Eq:Regula-result2}
C_c(\psi,\epsilon)=\max_{M\in B^*(\psi,\epsilon)} S_{\min}(\Delta(M)),
\end{equation}
where $B^*(\psi,\epsilon)=\{M:0\leq M\leq \mbb{I}, \;F^2(\psi,M)>1-\epsilon\}$ is the so-called operator ball around $\psi$.  Note that $B^*(\psi,\epsilon)\supset b(\psi,\epsilon)\supset b_*(\psi,\epsilon)$.  Our lower bound in Theorem \ref{Thm:pure_state} therefore implies that the maximum in Eq. \eqref{Eq:Regula-result2} is attained by a pure state $M$.

%However the converse for our result does not depend on the exact characterization of the incoherent operation and hence applies to all incoherent operations, while the achievablity holds for MIO, IO and SIO because of the inclusion relationship SIO $\subseteq $ IO $\subseteq $ MIO. 

\section{Coherence concentration for an ensemble of pure states} \label{ensemble_section}
For any given pure state ensemble $\mathfrak{E} = \lbrace p_i, \psi_i \rbrace_i$ we define the coherence concentration for $\mathfrak{E}$ as  :
\begin{equation}
C_{c}(\mathfrak{E}, \epsilon) := \max_{M \in \mathbb{N}} \left\lbrace \log_2M : \max_{ \lbrace\Lambda_{i}\rbrace_i}  F^{2}\left(\sum_{i}p_i\Lambda_i(\psi_i), \Phi_{M} \right) \geq 1 - \epsilon \right\rbrace,
\end{equation}
where $\Lambda_i$ are incoherent operators. 
The one-shot coherence of assistance is then given by,
\begin{equation}
C_{a}(\rho, \epsilon) = \max_{\mathfrak{E}}C_{c}(\mathfrak{E}, \epsilon),
\end{equation}
where $\mathfrak{E}$ are all possible pure state ensemble decompositions of $\rho$. We will now define for any ensemble $\mathfrak{E} = \lbrace p_i, \psi_i \rbrace_i$ the following quantity :
\begin{equation}
F_{min}^{\Delta}(\mathfrak{E}) := \min_i S_{min}(\Delta(\psi_i)).
\end{equation}
This is an estimate of the minimum coherence that can be distilled from the ensemble $\mathfrak{E}$. Also for any ensemble $\mathfrak{E}$ and $\epsilon \geq 0$
 we define the set :
 
 \begin{equation}
 b(\mathfrak{E}, \epsilon) := \left\lbrace  \overline{\mathfrak{E}} = \lbrace p_i, \overline{\psi}_i \rbrace_i : \sum\limits_i p_i F^2(\overline{\psi}_i, \psi_i) \geq 1 - \epsilon \right\rbrace.
 \end{equation}
Now we state our main result :

\vspace{0.2cm}
\noindent\fbox{\begin{minipage}{\dimexpr\textwidth-2\fboxsep-2\fboxrule\relax}
\centering
\begin{theorem}\label{distilable_ensemble_theorem}
For  any given ensemble  $\mathfrak{E} = \lbrace p_i, \psi_i \rbrace_i$ of pure states, and any $\epsilon \geq 0$,
\begin{equation}
\max_{\overline{\mathfrak{E}} \in b(\mathfrak{E}, \epsilon)}F_{min}^{\Delta}(\overline{\mathfrak{E}}) - \delta \leq C_{c}(\mathfrak{E}, \epsilon) \leq \max_{\overline{\mathfrak{E}} \in b(\mathfrak{E}, 2\epsilon)} F_{min}^{\Delta}(\overline{\mathfrak{E}}),
\end{equation}
where $0 \leq \delta \leq 1 $ is a number to ensure that the lower limit is the logarithm of an integer.
\end{theorem}
\end{minipage}}
\vspace{0.2cm}

Our proof of Theorem \ref{distilable_ensemble_theorem} follows in parallel to the proof of Theorem \ref{Thm:pure_state}.  For the lower bound, let $\overline{\mathfrak{E}} = \lbrace p_i, \overline{\psi}_i \rbrace_i $ be any ensemble such that $\overline{\mathfrak{E}}\in b(\mathfrak{E},\epsilon)$, i.e. $\sum\limits_{i} p_iF^2(\psi_i, \overline{\psi_i}) \geq 1  - \epsilon$.  As in the proof of Theorem \ref{Thm:pure_state}, we know that for each pure state $\psi_i$ Bob can distill a maximally coherent state of length $\log_2 \left\lfloor 2^{S_{min}(\Delta(\psi_i))}  \right\rfloor $ without error.    Then there exists  a set of incoherent operations $\lbrace \Lambda_i \rbrace_i $ such that $\Lambda_i(\overline{\psi}_i)=\Phi_{M(\mathfrak{\overline{E}})}$, where $M(\mathfrak{\overline{E}})\equiv \min\limits_i \left\lfloor 2^{S_{min}(\Delta(\psi_i))}  \right\rfloor $.  This is because each $\overline{\psi}_i \in \overline{\mathfrak{E}}$ can attain a maximally coherent state of at least length $M(\overline{\mathfrak{E}})$ using incoherent operations. Then,
\begin{equation}
\begin{split}
1 - \epsilon &\leq \sum_{i}p_i F^2(\psi_i, \overline{\psi}_i), \\
&\leq \sum_{i}p_iF^2(\Lambda_i(\psi_i), \Lambda_i(\overline{\psi}_i)), \\
&= \sum_{i}p_iF^2(\Lambda_i(\psi_i), \Phi_{M(\overline{\mathfrak{E}})}) ,\\
&= F^2\left(\sum\limits_i p_i \Lambda_i\left(\psi_i\right), \Phi_{M(\overline{\mathfrak{E}})}\right),
\end{split}
\end{equation}
where the second line follows from the monotonicity of fidelity under CP maps.  Since this holds for any $\overline{\mathfrak{E}}\in b(\mathfrak{E},\epsilon)$, we conclude that
\begin{equation}
\begin{split}
C_{c}(\mathfrak{E}, \epsilon) & \geq \max_{\overline{\mathfrak{E}} \in b(\mathfrak{E}, \epsilon)} \min_i S_{min}(\Delta(\psi_i)) - \delta  ,\\
&= \max_{\overline{\mathfrak{E}} \in b(\mathfrak{E}, \epsilon)} F_{min}^{\Delta}(\mathfrak{E}) - \delta,
\end{split}
\end{equation}
thus proving the lemma.

For the converse, suppose that $C_c(\mf{C},\epsilon)=\log_2M$.  Then there exists a family of incoherent maps $\{\Lambda_i\}_i$ such that
\begin{align}
1-\epsilon\leq F^2\left(\sum_i p_i\Lambda_i(\psi_i),\Phi_M \right) = \sum_i p_i\bra{\Phi_M}\Lambda_i(\psi_i)\ket{\Phi_M}.
\end{align}
Since each $\Lambda_i$ is incoherent, for any $\gamma\in\mc{I}$ we have that 
\begin{equation}
\Phi_M\Lambda_i(\gamma)\Phi_M\leq \frac{1}{M}\Phi_M.
\end{equation} 
With $\Lambda_i(\psi_i)\leq\mbb{I}$, we can multiply both sides of the previous equation by $\Lambda_i(\psi_i)$ and take the trace to obtain
\begin{align}
\label{Eq:MainIneq}
\log_2 M&\leq -\log\tr\left[\Phi_M\Lambda_i(\psi_i)\Phi_M\Lambda_i(\gamma)\right]\notag\\
&=-\log\tr\left[\Lambda^*_i\left(\Phi_M\Lambda_i(\psi_i)\Phi_M\right)\gamma\right]\notag\\
&=-\log\tr[Q_i\gamma]\notag\\
&\leq -\log\tr\left[\sqrt{Q_i}\psi_i\sqrt{Q_i}\gamma\right]\notag\\
&\leq -\log\tr[\overline{\psi}_i\gamma],
\end{align}
where we have used the fact that $\sqrt{Q} \psi_i \sqrt{Q} \leq Q$ and we have introduced the normalized state $\ket{\overline{\psi}_i} \equiv\frac{\sqrt{Q_i} \ket{\psi_i}}{\sqrt{\bra{\psi_i} Q_i\ket{\psi_i}}}$.  Define the pure state ensemble $\overline{\mf{E}}\equiv\{\ket{\overline{\psi}_i},p_i\}_i$.  Returning to equation \eqref{Eq:MainIneq}, we can choose the incoherent state $\gamma$ to an eigenvector associated with the largest eigenvalue of $\Delta(\overline{\psi}_i)$.  Using this inequality on every $\ket{\overline{\psi}_i}\in\overline{\mf{E}}$, we obtain 
\begin{equation}
\log_2 M\leq \min_i S_{\min}(\Delta(\overline{\psi}_i))=F^\Delta_{\min}(\overline{\mf{C}}).
\end{equation}
It remains to show that $\overline{\mf{E}}\in b(\mf{E},2\epsilon)$.  Using the inequality in Eq. \eqref{Eq:Fidelity-pure-approx}, we have
\begin{align}
\sqrt{\sum_i p_i F^2(\psi_i,\overline{\psi}_i)}&\geq\sqrt{\sum_{j}p_i\tr[Q_i\psi_i]}\notag\\
&=\sqrt{\sum_i p_i\bra{\Phi_M}\Lambda_i(\psi_i)\ket{\Phi_M}^2}\notag\\
&\geq \sum_ip_i\bra{\Phi_M}\Lambda_i(\psi_i)\ket{\Phi_M}\geq 1-\epsilon,
\end{align}
where the second inequality follows from the concavity of the function $f(x)=\sqrt{x}$.  Hence $\sum_i p_i F^2(\psi_i,\overline{\psi}_i)\geq (1-\epsilon)^2\geq 1-2\epsilon$.

\section{Asymptotic coherence of assistance }\label{asymp_section}
For a mixed state $\rho\equiv\rho^B$, its one-shot coherence of assistance is given by
\begin{equation}
C_a(\rho, \epsilon) = \max_{\mathfrak{E}} C_{c}(\mathfrak{E}, \epsilon) ,
\end{equation}
where the maximization is over all ensemble decompositions $\mathfrak{E}$ of $\rho$.
The coherence of assistance for $\rho$ is defined by
\begin{equation}
D_a(\rho) = \max_{\mf{E}=\lbrace p_i, \psi_i \rbrace_i} \sum\limits_i p_i S(\Delta(\psi_i)),
\end{equation}
with its regularized version being $D_a^{\infty}(\rho) = \lim\limits_{n \rightarrow \infty}\frac{1}{n} D_{a}(\rho^{\otimes n} )$. The asymptotic assisted coherence concentration for Alice and Bob sharing a pure state $\ket{\psi}^{AB}$ is given by \cite{chitambar_2016_assisted},
\begin{equation} 
D_{c}^{A | B}(\ket{\psi}^{AB}) = D_a^{\infty}(\rho^B) = S(\Delta(\rho^B)),
\end{equation}
where $\rho^{B} = \Tr_A(\ket{\psi}^{AB})$. We define the asymptotic limit of the one-shot coherence of assistance as,
\begin{equation}
C_{a}^{\infty}(\rho) = \lim\limits_{\epsilon \rightarrow 0}\lim\limits_{n \rightarrow \infty}\frac{1}{n} C_a(\rho^{\otimes n}, \epsilon) .
\end{equation}
We will now show that under this limit we recover the asymptotic expression.

\vspace{0.2cm}
\noindent\fbox{\begin{minipage}{\dimexpr\textwidth-2\fboxsep-2\fboxrule\relax}
\centering
\begin{theorem}\label{asyp_assist_theorem}
For any state $\rho$,
\begin{equation}
C_{a}^{\infty}(\rho)  = D_a^{\infty}(\rho).
\end{equation}
\end{theorem}
\end{minipage}}
\vspace{0.2cm}

\begin{lemma} \label{lemma:asy_proof_1}
For any state $\rho$,
\begin{equation}
C_a^\infty(\rho)\leq D_a^\infty(\rho).
\end{equation}
\end{lemma}
Proof : 
Suppose $\rho$ has support on a $d$-dimensional space.  From Theorem \ref{distilable_ensemble_theorem}, we have
\begin{align}
C_a(\rho^{\otimes n},\epsilon)\leq\max_{\mf{E}}\max_{\overline{\mf{E}}\in b(\mf{E},2\epsilon)}F^{\Delta}_{\min}(\overline{\mf{E}})&\leq \max_{\mf{E}}\max_{\overline{\mf{E}}\in b(\mf{E},2\epsilon)}\sum_i p_iS_{\min}(\Delta(\overline{\psi}_i))\notag\\
&\leq \max_{\mf{E}}\max_{\overline{\mf{E}}\in b(\mf{E},2\epsilon)}\sum_i p_iS(\Delta(\overline{\psi}_i)),
\end{align}
where the first maximization is taken over all ensembles $\mf{E}$ generating $\rho^{\otimes n}$.  To bound the last term introduce the QC states $\sigma^{BX}=\sum_ip_i\psi_i\otimes\op{i}{i}$, $\overline{\sigma}^{BX}=\sum_i p_i\overline{\psi}_i\otimes\op{i}{i}$, and note
\begin{align}
||\sigma^{BX}-\overline{\sigma}^{BX}||&=\sum_i p_i||\psi_i-\overline{\psi}_i||=2\sum_i p_i\sqrt{1-F^2(\psi_i,\overline{\psi}_i)}\leq 2 \sqrt{1-\sum_ip_iF^2(\psi_i,\overline{\psi}_i)}\leq 2\sqrt{2\epsilon}.
\end{align}
If we let $\Delta^B$ denote the dephasing map on system $B$ then we further have $\delta:=||\Delta^B(\sigma^{BX})-\Delta^B(\overline{\sigma}^{BX})||\leq 2\sqrt{2\epsilon}$.  An application of the Alicki-Fannes inequality \cite{Alicki_2004_continuity} to the (classical) states $\Delta^B(\sigma^{BX})$ and $\Delta^B(\overline{\sigma}^{BX})$ yields
\begin{equation}
\left|\sum_{i}p_iS(\Delta(\overline{\psi}_i))-\sum_{i}p_iS(\Delta(\psi_i))\right|\leq 4\delta n\log(d)+h(\delta),
\end{equation}
where $h(\delta) := -\delta\log_2(\delta) - (1 - \delta) \log_2 (1 - \delta)$, is the binary entropy function. Hence
\begin{align}
C_a(\rho^{\otimes n},\epsilon)&\leq\max_{\mf{E}}\sum_{i}p_iS(\Delta(\psi_i))+4\delta n\log(d)+h(\delta)=D_a(\rho^{\otimes n})+4\delta n\log(d)+h(\delta).
\end{align}
Dividing both sides by $n$ and taking the limits $n\to\infty$, $\epsilon\to 0$ yields the desired result.

\begin{definition}
We define the quantum-classical state corresponding to any pure state ensemble $\mathfrak{E} = \lbrace p_i, \psi_i^B \rbrace_i$ as,
\begin{equation}
\sigma_{\mathfrak{E}}^{BZ} := \sum\limits_i p_i \psi_i^B \otimes \pi_i^Z.
\end{equation}
where $\pi_i^Z$ are orthogonal rank one projectors $\ketbra{i}{i}^Z$. 
\end{definition}
We define the function $\overline{C}_{min} :\mc{D}(\mc{H}^B \otimes \mc{H}^Z) \rightarrow \mbb{R} $ which is a smoothed version of $C_{min}(\cdot)$ introduced in equation \eqref{Eq:C_min_def_eq} but defined for Q.C. states;
\begin{equation}
\overline{C}_{min}^{\epsilon}(\sigma^{BZ}_{\mathfrak{E}}) := \max_{\overline{\mathfrak{E}} \in b(\mathfrak{E}, \epsilon)} \min_{\nu^{BZ} \in \mathcal{I}}  S_{0}(\overline{\sigma}^{BZ}_{\mathfrak{E}}  \| \nu^{BZ}).
\end{equation} 
Further we will make use of the following lemmas,
\begin{lemma}\label{C_a_bound_lemma}
For any state $\rho^B$ and any $\epsilon \geq 0$,
\begin{equation}
\max_{\mathfrak{E}}\overline{C}_{min}^{ \frac{\epsilon}{2}}(\sigma^{BZ}_{\mathfrak{E}}) - \delta \leq C_{a}(\rho^B, \epsilon) ,
\end{equation}
where the maximization is taken over all ensembles $\mathfrak{E} = \lbrace p_i, \psi_i \rbrace_i$ such that $\rho^B = \sum\limits_i p_i \psi_i$ , $\sigma^{BZ}_{\mathfrak{E}} := \sum\limits_i p_i \psi_i \otimes \pi_i $ and $0 \leq \delta \leq 1$ ensures the lower limit is the logarithm of a positive integer.
\end{lemma}
Proof: Notice that,
\begin{equation}
\begin{split}
\overline{C}_{min}^{\frac{\epsilon}{2}}(\sigma^{BZ}_{\mathfrak{E}}) &:= \max_{\overline{\mathfrak{E}} \in b(\mathfrak{E}, \frac{\epsilon}{2})}\min_{\nu^{BZ} \in \mathcal{I}} \left\lbrace -\log_2\Tr\left( \Pi_{\sigma^{BZ}_{\overline{\mathfrak{E}}} } \nu^{BZ} \right)  \right\rbrace ,\\
&= \max_{ \lbrace p_i, \overline{\phi}_i \rbrace_i \in b(\mathfrak{E}, \frac{\epsilon}{2})} \min_i \min_{\nu^{B} \in \mathcal{I}}\left\lbrace -\log_2 \Tr (\overline{\phi}_i \nu^B) \right\rbrace ,\\
&= \max_{ \lbrace p_i, \overline{\phi}_i \rbrace_i \in b(\mathfrak{E}, \frac{\epsilon}{2})} \min_i \left\lbrace -\log_2 \lambda_{max}(\Delta(\overline{\phi}_i))\right\rbrace ,\\
&= \max_{ \lbrace p_i,  \overline{\phi}_i \rbrace_i \in b(\mathfrak{E}, \frac{\epsilon}{2})} \min_i S_{min}(\Delta(\overline{\phi}_i)) ,\\
&= \max_{  \mathfrak{\overline{E}} \in b(\mathfrak{E}, \frac{\epsilon}{2})} F_{min}^{\Delta}(\overline{\mathfrak{E}}), \\
&\leq C_{c}(\mathfrak{E}, \epsilon),
\end{split}
\end{equation}
where the inequality comes from theorem \ref{distilable_ensemble_theorem}. Maximizing over $\mathfrak{E}$ proves the lemma. 
\begin{lemma}\label{C_qc_C_r_lemma}
Given a quantum classical state $(\sigma^{BZ}_{\mathfrak{E}})^{\otimes n}$ and any general pure state ensemble $\mathfrak{E}_n = \lbrace p_{i}^{(n)}, \psi_i^n   \rbrace_i $ such that $(\sigma^{BZ}_{\mathfrak{E}})^{\otimes n} =  \sum\limits_i p_i^{(n)} \psi_i^n $, we have
\begin{equation}
\lim\limits_{\epsilon \rightarrow 0} \lim\limits_{n \rightarrow \infty}\frac{1}{n} \max_{\mathfrak{E}_n}\overline{C}_{min}^{\epsilon}(\sigma^{B^nZ^n}_{\mathfrak{E}_n}) \geq \max_{\mathfrak{E}}C_{r}(\sigma^{BZ}_{\mathfrak{E}}) ,
\end{equation}
where $C_r(\sigma)$ is relative entropy of coherence.
\end{lemma}
Proof: We need to use some results from the quantum information spectrum approach. 
\begin{definition}
Given a sequence of states $\hat{\rho} = \lbrace \rho^n\rbrace_{n = 1}^{\infty}$ with $\rho^n \in \mathcal{D}(\mathcal{H}^{\otimes n} )$ (set of density operators in $\mathcal{H}^{\otimes n}$) and positive operators $\hat{\sigma} = \lbrace \sigma^n\rbrace_{n = 1}^{\infty} $ with $\sigma^n \in \mathcal{B}(\mathcal{H}^{\otimes n}) $ (set of positive operators acting on $\mathcal{H}^{\otimes n}$), and defining $\Delta^n(\gamma) := \rho^n - 2^{n\gamma}\sigma^n$, the quantum spectral inf-divergence rate is defined as,

\begin{equation}\label{inf_spec_def_eqn}
\underline{D}(\hat{\rho} \| \hat{\sigma}) := \sup \left\lbrace \gamma : \liminf_{n \rightarrow \infty} \Tr \left(  \lbrace \Delta^n \geq 0 \rbrace \Delta^n \right) = 1 \right\rbrace ,
\end{equation}
where $\lbrace X \geq 0 \rbrace$ for a self-adjoint operator $X$ denotes the projector unto the non-negative eigenspace of $X$.
\end{definition}

\begin{lemma}\label{D-gamma-lemma}
Given a state $\rho_n$ and a self-adjoint operator $\omega_n$, for any real $\gamma$,we have,
\begin{equation}
\Tr\left(\lbrace \rho_n - 2^{n\gamma}\omega_n \rbrace \omega_n \right) \leq 2^{-n\gamma} .
\end{equation}
\end{lemma}
Proof : see \cite{datta_2009_min}.
\begin{lemma}\label{D_inf_ubound_lemma}
For any given state $\rho^B$, let $\mathfrak{E} = \lbrace p_i, \psi_i \rbrace$ denote a pure state decomposition and $\mathfrak{E}_n = \lbrace p_{i, n}, \psi_i^n \rbrace $ denote a pure state decomposition of the state $(\rho^{B}) ^{\otimes n}$, then we have,
\begin{equation}\label{inf_spec_lbound_eqn}
\lim\limits_{\epsilon \rightarrow 0} \lim\limits_{n \rightarrow \infty}\frac{1}{n} \max_{\mathfrak{E}_n}\overline{C}_{min}^{\epsilon}(\sigma^{B^nZ^n}_{\mathfrak{E}_n}) \geq \max_{\mathfrak{E}}\min_{\nu^{BZ} \in \mathcal{I}} \underline{D}(\hat{\sigma}_{\mathfrak{E}}^{BZ} \| \hat{\nu}^{BZ}),
\end{equation}
where $\hat{\sigma}_{\mathfrak{E}}^{BZ} = \left\lbrace (\sigma^{BZ}_{\mathfrak{E}})^{\otimes n} \right\rbrace_{n \geq 1}$ and $\hat{\nu}^{BZ} = \lbrace (\nu^{BZ})^{\otimes n} \rbrace_{n \geq 1}$.
\end{lemma}
Proof: Let $\mathfrak{E^*}$ be an ensemble such that it achieves the maximum in equation \eqref{inf_spec_lbound_eqn}. By definition we have,
\begin{equation}\label{equation_step}
\begin{split}
\max_{\mathfrak{E_n}}\overline{C}^{\epsilon}_{min} (\sigma^{B^nZ^n}_{\mathfrak{E_n}}) &= \max_{\mathfrak{E}_n} \max_{\overline{\mathfrak{E}}_n \in b(\mathfrak{E}_n, \epsilon)} \min_{\nu^{B^nZ^n} \in \mathcal{I}} S_{0}(\overline{\sigma}^{B^nZ^n}_{\overline{\mathfrak{E}}_n} \| \nu^{B^nZ^n}), \\
&\geq  \max_{\mathfrak{E}} \max_{\overline{\mathfrak{E}}_n \in b(\mathfrak{E}^{\otimes n}, \epsilon)} \min_{\nu^{B^nZ^n} \in \mathcal{I}} S_{0}(\overline{\sigma}^{B^nZ^n}_{\overline{\mathfrak{E}}_n} \| \nu^{B^nZ^n}), \\
&\geq \max_{\overline{\mathfrak{E}}_n \in b((\mathfrak{E}^*)^{\otimes n}, \epsilon)} \min_{\nu^{B^nZ^n} \in \mathcal{I}} S_{0}(\overline{\sigma}^{B^nZ^n}_{\overline{\mathfrak{E}}_n} \| \nu^{B^nZ^n}),
\end{split}
\end{equation}
where $\mathfrak{E}^{\otimes n}$ is the product  pure state ensemble $\lbrace p_i, \psi_i \rbrace^{\otimes n}$ .
For each $\nu^{B^nZ^n}$ and any $\gamma \in \mathbb{R}$ we define the projector,
\begin{equation}
P^n_{\gamma} \equiv P_{\gamma}^n(\nu^{B^nZ^n}) := \lbrace (\sigma_{\mathfrak{E^*}}^{BZ} )^{\otimes n} - 2^{n \gamma} \nu^{B^nZ^n} \geq 0\rbrace .
\end{equation}
Since $\nu^{B^nZ^n}$ are incoherent states, the projector $P_{\gamma}^{n}$ also has a Q.C. structure. Let $\hat{\sigma}^{BZ}_{\mathfrak{E^*}}$ be the i.i.d. sequence of states $\lbrace ( \sigma^{BZ}_{\mathfrak{E}^*} )^{\otimes n}\rbrace_{n =1}^{\infty}$. For a sequence $\hat{\nu}^{BZ} := \lbrace \nu^{B^nZ^n}_{n} \rbrace_{n = 1}^{\infty}$ fix $\delta > 0$ and choose $\gamma \equiv \gamma(\hat{\nu}^{BZ}) := \underline{D}(\hat{\sigma}^{BZ}_{\mathfrak{E^*}} \| \hat{\nu}^{BZ}_n) - \delta$. Then from the definition of the quantum inf-divergence rate in equation \eqref{inf_spec_def_eqn}, there exists an $n$ large enough such that,
\begin{equation}
\Tr \left( P_{\gamma}^n (\sigma_{\mathfrak{E}^*}^{BZ})^{\otimes n} \right) \geq 1 - \epsilon ,
\end{equation}
for any $\epsilon \geq 0$. Here we have used the fact that the quantum inf-divergence rate can be arternatively defined as (see prop 2. in  \cite{bowen_2006_beyond} )
\begin{equation}
\underline{D}(\hat{\rho} \| \hat{\sigma}) := \sup \left\lbrace \gamma : \liminf_{n \rightarrow \infty} \Tr \left(  \lbrace \Delta^n \geq 0 \rbrace \rho^n \right) = 1 \right\rbrace.
\end{equation}
Now we define,
\begin{equation}
\frac{P_{\gamma}^n (\sigma^{BZ}_{\mathfrak{E^*}})^{\otimes n} P_{\gamma}^n}{\tr\left(P^n_{\gamma} (\sigma^{BZ}_{\mathfrak{E^*}})^{\otimes n} \right)}  = \frac{\sum_i p_{i, n} \overline{\psi}_i^n \otimes \pi_{i}^n}{\tr(P^n_{\gamma}(\sigma^{BZ}_{\mathfrak{E}^*})^{\otimes n})}  \equiv \omega^{B^nZ^n}_{\mathfrak{E}_n^{\prime}, \gamma}(\nu^{B^nZ^n}) =: \omega^{B^nZ^n}_{\mathfrak{E}_n^{\prime}, \gamma},
\end{equation}
where $\pi_i^n = \op{i^n}{i^n}$ and  $\mathfrak{E}_n^{\prime}$ is the pure state ensemble $\lbrace p_{i, n}, \frac{\overline{\psi}_i^n}{\tr(P_{\gamma}^n (\sigma_{\mathfrak{E}^*}^{BZ})^{\otimes n} )} \rbrace_i$ with $\overline{\psi}^n_i = \Tr_{Z^n}( P^n_{\gamma} ( \psi_i^n \otimes \pi_i^n) P^n_{\gamma} ) $. We will now show that $\mathfrak{E}^{\prime}_n \in b((\mathfrak{E}^*)^{\otimes n }, \epsilon)$. Since $P_{\gamma}^n$ has a Q.C. structure, we can write it as $P_{\gamma}^n = \sum_i \Pi_{\gamma, i}^{n} \otimes \pi_i^n $. Where $\Pi_{\gamma, i}^n$ are projectors acting on the Hilbert space $ (\mc{H }^{B})^{\otimes n}$. Now we have,

\begin{equation}\label{Eq:P_sigma_bound}
\begin{split}
1 - \epsilon &\leq \Tr \left( P_{\gamma}^n (\sigma_{\mathfrak{E}^*}^{BZ})^{\otimes n} \right) ,\\
&= \tr\left( \sum\limits_i p_{i, n}\Pi_{\gamma, i}^n \psi_i^n  \otimes \pi_i^n   \right) , \\
&= \sum\limits_i p_{i, n}\tr\left( \Pi_{\gamma, i}^n \psi_i^n\right) .\\
\end{split}
\end{equation}
but note that,
\begin{equation}
\begin{split}
F \left( \frac{\overline{\psi}_i^n}{\tr(P_{\gamma}^n (\sigma_{\mathfrak{E}^*}^{BZ})^{\otimes n})}, \psi_i^n \right) &= \frac{1}{\sqrt{\sum\limits_j p_{j, n} \tr(\Pi_{\gamma, j}^n \psi_i^n)}}\Tr \left( \sqrt{\braket{\psi_i^n | \Pi_{\gamma, i}^n \op{\psi_i^n}{\psi_i^n} \Pi_{\gamma, i}^n | \psi_i^n  }  } \right) ,\\
&= \frac{1}{\sqrt{\sum\limits_j p_{j, n} \tr(\Pi_{\gamma, j}^n \psi_i^n)}} \tr\left( \Pi_{\gamma, i}^n \psi_i^n\right).
\end{split}
\end{equation}
Hence we have,
\begin{equation}\label{Eq:fid_ensemble_eqn}
\begin{split}
\sum\limits_i p_{i, n} F\left( \frac{\overline{\psi}_i^n}{\tr(P_{\gamma}^n (\sigma_{\mathfrak{E}^*}^{BZ})^{\otimes n} )}, \psi_i^n \right) &= \sqrt{\sum\limits_j p_{j, n} \tr(\Pi_{\gamma, j}^n \psi_i^n)} ,\\
& \geq \sum\limits_j p_{j, n} \tr(\Pi_{\gamma, j}^n \psi_i^n) ,\\
&\geq 1 - \epsilon,
\end{split}
\end{equation}
where the last inequality follows from equation \eqref{Eq:P_sigma_bound}. Equation \eqref{Eq:fid_ensemble_eqn} implies that $\mathfrak{E}_n^{\prime} \in S_{= }(\mathfrak{(E^*)^{\otimes n}}, \epsilon)$. Proceeding from equation \eqref{equation_step} we have
\begin{equation}
\begin{split}
&\lim\limits_{n \rightarrow \infty} \frac{1}{n}  \left( \max_{\overline{\mathfrak{E}}_n \in b( (\mathfrak{E}^*)^{\otimes n}, \epsilon)} \min_{\nu^{B^nZ^n} \in \mathcal{I}} S_{0}(\sigma^{B^nZ^n}_{\overline{\mathfrak{E}}_n} \| \nu_n^{B^nZ^n}) \right) ,\\
&\geq \lim\limits_{n \rightarrow \infty} \frac{1}{n} \min_{\nu^{B^nZ^n} \in \mathcal{I}} S_{0}(\omega^{B^nZ^n}_{\mathfrak{E}_n^{\prime}, \gamma} \| \nu_n^{B^nZ^n})  ,\\  
&= \lim\limits_{n \rightarrow \infty} \frac{1}{n} \min_{\nu^{B^nZ^n} \in \mathcal{I}} \left\lbrace -\log_2 \Tr\left( \Pi_{\omega^{B^nZ^n}_{\mathfrak{E}_n^{\prime}, \gamma}} \nu_n^{B^nZ^n}\right) \right\rbrace ,\\
&\geq \lim\limits_{n \rightarrow \infty} \frac{1}{n} \min_{\nu^{B^nZ^n} \in \mathcal{I}} \left\lbrace -\log_2 \Tr\left( P_{\gamma}^n \nu_n^{B^nZ^n}\right) \right\rbrace ,\\
&\geq \min_{\hat{\nu}^{BZ} } \gamma(\hat{\nu}^{BZ}) ,\\
&= \underline{D}(\hat{\sigma}^{BZ}_{\mathfrak{E^*}} \| \hat{\nu}^{BZ}_n) - \delta ,\\
&= \max_{\mathfrak{E}} \underline{D}(\hat{\sigma}^{BZ}_{\mathfrak{E}} \| \hat{\nu}^{BZ}_n) - \delta .
\end{split}
\end{equation}
For the second inequality, we have used the fact that $\Pi_{\omega_{\mathfrak{E}_n^{\prime}, \gamma}^{B^nZ^n}} \leq P^n_{\gamma}$ and the third inequality follows from lemma \ref{D-gamma-lemma}. As this holds for arbitrary $\delta \geq 0$ we recover the statement of lemma \ref{D_inf_ubound_lemma} in the limit $\epsilon \rightarrow 0$.

\begin{lemma}\label{D_inf_c_r_lemma}
For any sequence of states $\hat{\rho} = \lbrace \rho^{\otimes n} \rbrace_{n \geq 1}$, 
\begin{equation}
 \min_{\hat{\sigma}} \underline{D}(\hat{\rho} \| \hat{\sigma}) = C_{r}(\rho),
 \end{equation} 
 where $\hat{\sigma} = \lbrace \sigma^n \rbrace_{n \geq 1}$ with $\sigma^n \in \mathcal{I}$ and $ C_{r}(\rho) =  \min\limits_{\delta \in \mathcal{I}} S(\rho \| \delta)$ is the relative entropy of coherence.
\end{lemma}
Proof: Consider the family of sets $\mathcal{M} := \lbrace \mathcal{M}_n \rbrace_{n \geq 1}$
\begin{equation}
\mathcal{M}_n := \lbrace \delta_n \in \mathcal{I}_n  \rbrace_{n \geq 1}
\end{equation}
where $\mathcal{I}_n$ is the set of incoherent states in $\mathcal{H}^{\otimes n}$. 

\begin{prop}\label{stein_prop}
The family of sets $\mathcal{M}$ satisfies the conditions required to apply the generalized Stein's lemma (proposition III.1 in \cite{brandao_2010_generalization}) .
\end{prop}
Proof: see appendix \ref{stein_appendix}.

From proposition \ref{stein_prop} we have for a given state $\rho$,
\begin{equation}
\mathcal{S}_{\mathcal{M}}^{\infty} := \frac{1}{n} \mathcal{S}_{\mathcal{M}_n}(\rho^{\otimes n}),
\end{equation}
with $\mathcal{S}_{\mathcal{M}_n}(\rho^{\otimes n}) := \min\limits_{\delta_n \in \mathcal{M}_n} S(\rho^{\otimes n} \| \delta_n)$. Let $\Delta_n(\gamma) = \rho^{\otimes n} - 2^{n \gamma}\delta_n$. Then from the generalized Stein's lemma in \cite{brandao_2010_generalization} it follows that for $\gamma > \mathcal{S}_{\mathcal{M}}^{\infty}(\rho)$,
\begin{equation}
\lim\limits_{n \rightarrow \infty} \min_{\delta_n \in \mathcal{M}_n} \Tr \left( \lbrace \Delta_n(\gamma) \geq 0 \rbrace \Delta_n \right) = 0.
\end{equation}
This implies that $\min\limits_{\hat{\sigma}}\underline{D}(\hat{\rho} \| \hat{\sigma}) \leq S_{\mathcal{M}}^{\infty}(\rho)$. Conversely, for $\gamma < \mathcal{S}_{\mathcal{M}}^{\infty}(\rho)$,
\begin{equation}
\lim\limits_{n \rightarrow \infty} \min_{\delta_n \in \mathcal{M}_n} \Tr \left( \lbrace \Delta_n(\gamma) \geq 0 \rbrace \Delta_n \right) = 1,
\end{equation}
which implies that $\min\limits_{\hat{\sigma}}\underline{D}(\hat{\rho} \| \hat{\sigma}) \geq \mathcal{S}_{\mathcal{M}}^{\infty}(\rho) $. Thus we have,
\begin{equation}
\underline{D}(\hat{\rho} \| \hat{\sigma}) = \mathcal{S}_{\mathcal{M}}^{\infty}(\rho).
\end{equation}
But by definition $\mathcal{S}_{\mathcal{M}}^{\infty}(\rho) \equiv C_r^{\infty}(\rho) := \lim\limits_{n \rightarrow \infty} \frac{1}{n} \min\limits_{\delta_n \in \mathcal{I}} S(\rho^{\otimes n} \| \delta_n) = C_r(\rho)$ because of the additivity of the relative entropy of coherence \cite{winter_2016_operational}, thus proving lemma \ref{D_inf_c_r_lemma}. Lemma \ref{D_inf_ubound_lemma} and lemma \ref{D_inf_c_r_lemma} together prove lemma \ref{C_qc_C_r_lemma}.

\begin{lemma}\label{asy_th_1_lemma}
For any bipartite state $\rho^B$,
\begin{equation}
C_{a}^{\infty}(\rho^B) \geq \lim_{n \rightarrow \infty} \frac{1}{n} D_a((\rho^B)^{\otimes n}) \equiv D_a\infty(\rho).
\end{equation}
\end{lemma}
Proof: Let $\mathfrak{E} = \lbrace p_i, \psi_i \rbrace_i$ be a pure state ensemble decomposition of $\rho$ and $\mathfrak{E}_n = \lbrace p_{i^n}, \psi_{i^n}^{B^n} \rbrace_{i^n}$ be such a decomposition of $(\rho^B)^{\otimes n}.$ As before we define the q.c. state,
\begin{equation}
\sigma_{\mathfrak{E}_n}^{B^nZ^n} = \sum\limits_{i}p_{i^n}\phi_{i^n}^{B^n} \otimes \pi_{i^n}^{Z^n},
\end{equation}
where $\pi_{i^n}^{Z^n} = \ketbra{i^n}{i^n}$ is the incoherent basis in $\mathcal{H}_Z^{\otimes n}$. From lemma \ref{C_a_bound_lemma} we know that,
\begin{equation}
C_a((\rho^{B})^{\otimes n}, \epsilon) \geq \max_{\mathfrak{E}_n}\overline{C}_{min}^{\frac{\epsilon}{2}}(\sigma_{\mathfrak{E}_n}^{B^nZ^n}) - \delta_n,
\end{equation}
where $0 \leq \delta_n \leq 1$. So we have,
\begin{equation}
\begin{split}
C_a^{\infty}(\rho^B) &:= \lim\limits_{\epsilon \rightarrow 0} \lim\limits_{n \rightarrow \infty} \frac{1}{n} C_a((\rho^B)^{\otimes n}, \epsilon) ,\\
&\geq \lim\limits_{\epsilon \rightarrow 0} \lim\limits_{n \rightarrow \infty} \frac{1}{n} \max_{\mathfrak{E}_n} \overline{C}_{min}^{\frac{\epsilon}{2}}(\sigma_{\mathfrak{E}_n}^{B^nZ^n}) ,\\
&\geq \max_{\mathfrak{E}}C_{r}(\sigma^{BZ}_{\mathfrak{E}}), \\
\end{split} 
\end{equation}
where we have used lemma \ref{C_a_bound_lemma} for tne first inequality and lemma \ref{C_qc_C_r_lemma} for the last inequality. But since $\sigma_{\mathfrak{E}}^{BZ}$ is a quantum classical state, we have (see appendix \ref{eqn_proof_appendix})
\begin{equation}\label{qc_c_r_eqn}
C_r\left(\sum\limits_i p_i \phi_i^B \otimes \pi_i^Z \right ) = \sum\limits_i p_i C_r(\phi_i).
\end{equation}
Hence we have,
\begin{equation}
C_a^{\infty}(\rho^B) := \lim\limits_{\epsilon \rightarrow 0} \lim\limits_{n \rightarrow \infty} \frac{1}{n} C_a((\rho^B)^{\otimes n}, \epsilon) \geq \max_{\lbrace p_i, \phi_i^B \rbrace} \sum\limits_i p_i C_r(\phi_i^B) = D_a(\rho^B).
\end{equation}
Lemma \ref{lemma:asy_proof_1} and \ref{asy_th_1_lemma}  proves theorem \ref{asyp_assist_theorem}.

\section{Conclusions}\label{conclusion_section}
We have derived bounds for the one-shot concentration of maximally coherent states for pure states and average rate for an ensemble of pure states. Using this we have given bounds on the one-shot coherence of assistance and hence the assisted coherence concentration. We further show that asymptotically the one-shot quantity reduces to the correct known result. Finding the one-shot concentration rate for a more general scenario than assistance where communication is not restricted to being one-way and with multiple parties  helping Bob, the so called collaboration scenario, remains an open question. Our results highlight how techniques used in the resource theory of entanglement can find ready application to the resource theory of coherence and we hope it will help deepen understanding of the relationship between the two.

\section*{Acknowledgements}
We thank Bartosz Regula and Andreas Winter for valuable discussion. M.-H.~Hsieh was supported by an ARC Future Fellowship under Grant FT140100574 and by US Army Research Office for Basic Scientific Research Grant W911NF-17-1-0401.

\bibliographystyle{apsrev}
\bibliography{refs}

\begin{thebibliography}{33}
\expandafter\ifx\csname natexlab\endcsname\relax\def\natexlab#1{#1}\fi
\expandafter\ifx\csname bibnamefont\endcsname\relax
  \def\bibnamefont#1{#1}\fi
\expandafter\ifx\csname bibfnamefont\endcsname\relax
  \def\bibfnamefont#1{#1}\fi
\expandafter\ifx\csname citenamefont\endcsname\relax
  \def\citenamefont#1{#1}\fi
\expandafter\ifx\csname url\endcsname\relax
  \def\url#1{\texttt{#1}}\fi
\expandafter\ifx\csname urlprefix\endcsname\relax\def\urlprefix{URL }\fi
\providecommand{\bibinfo}[2]{#2}
\providecommand{\eprint}[2][]{\url{#2}}

\bibitem[{\citenamefont{Devetak et~al.}(2008)\citenamefont{Devetak, Harrow, and
  Winter}}]{devetak_2008_resource}
\bibinfo{author}{\bibfnamefont{I.}~\bibnamefont{Devetak}},
  \bibinfo{author}{\bibfnamefont{A.~W.} \bibnamefont{Harrow}},
  \bibnamefont{and} \bibinfo{author}{\bibfnamefont{A.~J.}
  \bibnamefont{Winter}}, \bibinfo{journal}{IEEE Transactions on Information
  Theory} \textbf{\bibinfo{volume}{54}}, \bibinfo{pages}{4587}
  (\bibinfo{year}{2008}).

\bibitem[{\citenamefont{Horodecki et~al.}(2009)\citenamefont{Horodecki,
  Horodecki, Horodecki, and Horodecki}}]{horodecki_2009_quantum}
\bibinfo{author}{\bibfnamefont{R.}~\bibnamefont{Horodecki}},
  \bibinfo{author}{\bibfnamefont{P.}~\bibnamefont{Horodecki}},
  \bibinfo{author}{\bibfnamefont{M.}~\bibnamefont{Horodecki}},
  \bibnamefont{and}
  \bibinfo{author}{\bibfnamefont{K.}~\bibnamefont{Horodecki}},
  \bibinfo{journal}{Reviews of modern physics} \textbf{\bibinfo{volume}{81}},
  \bibinfo{pages}{865} (\bibinfo{year}{2009}).

\bibitem[{\citenamefont{Plenio and Virmani}(2007)}]{plenio_2007_introduction}
\bibinfo{author}{\bibfnamefont{M.~B.} \bibnamefont{Plenio}} \bibnamefont{and}
  \bibinfo{author}{\bibfnamefont{S.}~\bibnamefont{Virmani}},
  \bibinfo{journal}{Quantum Info. Comput.} \textbf{\bibinfo{volume}{7}},
  \bibinfo{pages}{1} (\bibinfo{year}{2007}), ISSN \bibinfo{issn}{1533-7146}.

\bibitem[{\citenamefont{Gour et~al.}(2015)\citenamefont{Gour, M{\"u}ller,
  Narasimhachar, Spekkens, and Halpern}}]{gour_2015_resource}
\bibinfo{author}{\bibfnamefont{G.}~\bibnamefont{Gour}},
  \bibinfo{author}{\bibfnamefont{M.~P.} \bibnamefont{M{\"u}ller}},
  \bibinfo{author}{\bibfnamefont{V.}~\bibnamefont{Narasimhachar}},
  \bibinfo{author}{\bibfnamefont{R.~W.} \bibnamefont{Spekkens}},
  \bibnamefont{and} \bibinfo{author}{\bibfnamefont{N.~Y.}
  \bibnamefont{Halpern}}, \bibinfo{journal}{Physics Reports}
  \textbf{\bibinfo{volume}{583}}, \bibinfo{pages}{1} (\bibinfo{year}{2015}).

\bibitem[{\citenamefont{Brandao and
  Plenio}(2010)}]{brandao_2010_generalization}
\bibinfo{author}{\bibfnamefont{F.~G.} \bibnamefont{Brandao}} \bibnamefont{and}
  \bibinfo{author}{\bibfnamefont{M.~B.} \bibnamefont{Plenio}},
  \bibinfo{journal}{Communications in Mathematical Physics}
  \textbf{\bibinfo{volume}{295}}, \bibinfo{pages}{791} (\bibinfo{year}{2010}).

\bibitem[{\citenamefont{Gour and Spekkens}(2008)}]{gour_2008_resource}
\bibinfo{author}{\bibfnamefont{G.}~\bibnamefont{Gour}} \bibnamefont{and}
  \bibinfo{author}{\bibfnamefont{R.~W.} \bibnamefont{Spekkens}},
  \bibinfo{journal}{New Journal of Physics} \textbf{\bibinfo{volume}{10}},
  \bibinfo{pages}{033023} (\bibinfo{year}{2008}).

\bibitem[{\citenamefont{Horodecki et~al.}(2002)\citenamefont{Horodecki,
  Oppenheim, and Horodecki}}]{Horodecki-2002a}
\bibinfo{author}{\bibfnamefont{M.}~\bibnamefont{Horodecki}},
  \bibinfo{author}{\bibfnamefont{J.}~\bibnamefont{Oppenheim}},
  \bibnamefont{and}
  \bibinfo{author}{\bibfnamefont{R.}~\bibnamefont{Horodecki}},
  \bibinfo{journal}{Phys. Rev. Lett.} \textbf{\bibinfo{volume}{89}},
  \bibinfo{pages}{240403} (\bibinfo{year}{2002}).

\bibitem[{\citenamefont{Horodecki and Oppenheim}(2013)}]{Horodecki-2013a}
\bibinfo{author}{\bibfnamefont{M.}~\bibnamefont{Horodecki}} \bibnamefont{and}
  \bibinfo{author}{\bibfnamefont{J.}~\bibnamefont{Oppenheim}},
  \bibinfo{journal}{International Journal of Modern Physics B}
  \textbf{\bibinfo{volume}{27}}, \bibinfo{pages}{1345019}
  (\bibinfo{year}{2013}).

\bibitem[{\citenamefont{Brand\~ao and Gour}(2015)}]{Brandao-2015a}
\bibinfo{author}{\bibfnamefont{F.~G. S.~L.} \bibnamefont{Brand\~ao}}
  \bibnamefont{and} \bibinfo{author}{\bibfnamefont{G.}~\bibnamefont{Gour}},
  \bibinfo{journal}{Phys. Rev. Lett.} \textbf{\bibinfo{volume}{115}},
  \bibinfo{pages}{070503} (\bibinfo{year}{2015}).

\bibitem[{\citenamefont{Anshu et~al.}(2017)\citenamefont{Anshu, Hsieh, and
  Jain}}]{anshu_2017_quantifying}
\bibinfo{author}{\bibfnamefont{A.}~\bibnamefont{Anshu}},
  \bibinfo{author}{\bibfnamefont{M.-H.} \bibnamefont{Hsieh}}, \bibnamefont{and}
  \bibinfo{author}{\bibfnamefont{R.}~\bibnamefont{Jain}},
  \bibinfo{journal}{arXiv preprint arXiv:1708.00381}  (\bibinfo{year}{2017}).

\bibitem[{\citenamefont{Baumgratz et~al.}(2014)\citenamefont{Baumgratz, Cramer,
  and Plenio}}]{baumgratz_2014_quantifying}
\bibinfo{author}{\bibfnamefont{T.}~\bibnamefont{Baumgratz}},
  \bibinfo{author}{\bibfnamefont{M.}~\bibnamefont{Cramer}}, \bibnamefont{and}
  \bibinfo{author}{\bibfnamefont{M.}~\bibnamefont{Plenio}},
  \bibinfo{journal}{Physical review letters} \textbf{\bibinfo{volume}{113}},
  \bibinfo{pages}{140401} (\bibinfo{year}{2014}).

\bibitem[{\citenamefont{Aberg}(2006)}]{aberg_2006_quantifying}
\bibinfo{author}{\bibfnamefont{J.}~\bibnamefont{Aberg}},
  \bibinfo{journal}{arXiv preprint quant-ph/0612146}  (\bibinfo{year}{2006}).

\bibitem[{\citenamefont{Levi and Mintert}(2014)}]{levi_2014_quantitative}
\bibinfo{author}{\bibfnamefont{F.}~\bibnamefont{Levi}} \bibnamefont{and}
  \bibinfo{author}{\bibfnamefont{F.}~\bibnamefont{Mintert}},
  \bibinfo{journal}{New Journal of Physics} \textbf{\bibinfo{volume}{16}},
  \bibinfo{pages}{033007} (\bibinfo{year}{2014}).

\bibitem[{\citenamefont{Chitambar and Gour}(2016)}]{chitambar_2016_comparison}
\bibinfo{author}{\bibfnamefont{E.}~\bibnamefont{Chitambar}} \bibnamefont{and}
  \bibinfo{author}{\bibfnamefont{G.}~\bibnamefont{Gour}},
  \bibinfo{journal}{Physical Review A} \textbf{\bibinfo{volume}{94}},
  \bibinfo{pages}{052336} (\bibinfo{year}{2016}).

\bibitem[{\citenamefont{Du et~al.}(2015)\citenamefont{Du, Bai, and
  Guo}}]{du_2015_conditions}
\bibinfo{author}{\bibfnamefont{S.}~\bibnamefont{Du}},
  \bibinfo{author}{\bibfnamefont{Z.}~\bibnamefont{Bai}}, \bibnamefont{and}
  \bibinfo{author}{\bibfnamefont{Y.}~\bibnamefont{Guo}},
  \bibinfo{journal}{Physical Review A} \textbf{\bibinfo{volume}{91}},
  \bibinfo{pages}{052120} (\bibinfo{year}{2015}).

\bibitem[{\citenamefont{Winter and Yang}(2016)}]{winter_2016_operational}
\bibinfo{author}{\bibfnamefont{A.}~\bibnamefont{Winter}} \bibnamefont{and}
  \bibinfo{author}{\bibfnamefont{D.}~\bibnamefont{Yang}},
  \bibinfo{journal}{Physical Review Letters} \textbf{\bibinfo{volume}{116}},
  \bibinfo{pages}{120404} (\bibinfo{year}{2016}).

\bibitem[{\citenamefont{Yadin et~al.}(2016)\citenamefont{Yadin, Ma, Girolami,
  Gu, and Vedral}}]{yadin_2016_quantum}
\bibinfo{author}{\bibfnamefont{B.}~\bibnamefont{Yadin}},
  \bibinfo{author}{\bibfnamefont{J.}~\bibnamefont{Ma}},
  \bibinfo{author}{\bibfnamefont{D.}~\bibnamefont{Girolami}},
  \bibinfo{author}{\bibfnamefont{M.}~\bibnamefont{Gu}}, \bibnamefont{and}
  \bibinfo{author}{\bibfnamefont{V.}~\bibnamefont{Vedral}},
  \bibinfo{journal}{Physical Review X} \textbf{\bibinfo{volume}{6}},
  \bibinfo{pages}{041028} (\bibinfo{year}{2016}).

\bibitem[{\citenamefont{Streltsov et~al.}(2017)\citenamefont{Streltsov, Adesso,
  and Plenio}}]{streltsov_2017}
\bibinfo{author}{\bibfnamefont{A.}~\bibnamefont{Streltsov}},
  \bibinfo{author}{\bibfnamefont{G.}~\bibnamefont{Adesso}}, \bibnamefont{and}
  \bibinfo{author}{\bibfnamefont{M.~B.} \bibnamefont{Plenio}},
  \bibinfo{journal}{Reviews of Modern Physics} \textbf{\bibinfo{volume}{89}},
  \bibinfo{pages}{041003} (\bibinfo{year}{2017}).

\bibitem[{\citenamefont{Chitambar and Hsieh}(2016)}]{chitambar_2016_relating}
\bibinfo{author}{\bibfnamefont{E.}~\bibnamefont{Chitambar}} \bibnamefont{and}
  \bibinfo{author}{\bibfnamefont{M.-H.} \bibnamefont{Hsieh}},
  \bibinfo{journal}{Physical review letters} \textbf{\bibinfo{volume}{117}},
  \bibinfo{pages}{020402} (\bibinfo{year}{2016}).

\bibitem[{\citenamefont{Streltsov et~al.}(2015)\citenamefont{Streltsov, Singh,
  Dhar, Bera, and Adesso}}]{streltsov_2015_measuring}
\bibinfo{author}{\bibfnamefont{A.}~\bibnamefont{Streltsov}},
  \bibinfo{author}{\bibfnamefont{U.}~\bibnamefont{Singh}},
  \bibinfo{author}{\bibfnamefont{H.~S.} \bibnamefont{Dhar}},
  \bibinfo{author}{\bibfnamefont{M.~N.} \bibnamefont{Bera}}, \bibnamefont{and}
  \bibinfo{author}{\bibfnamefont{G.}~\bibnamefont{Adesso}},
  \bibinfo{journal}{Physical review letters} \textbf{\bibinfo{volume}{115}},
  \bibinfo{pages}{020403} (\bibinfo{year}{2015}).

\bibitem[{\citenamefont{Streltsov et~al.}(2016)\citenamefont{Streltsov,
  Chitambar, Rana, Bera, Winter, and Lewenstein}}]{streltsov_2016_entanglement}
\bibinfo{author}{\bibfnamefont{A.}~\bibnamefont{Streltsov}},
  \bibinfo{author}{\bibfnamefont{E.}~\bibnamefont{Chitambar}},
  \bibinfo{author}{\bibfnamefont{S.}~\bibnamefont{Rana}},
  \bibinfo{author}{\bibfnamefont{M.}~\bibnamefont{Bera}},
  \bibinfo{author}{\bibfnamefont{A.}~\bibnamefont{Winter}}, \bibnamefont{and}
  \bibinfo{author}{\bibfnamefont{M.}~\bibnamefont{Lewenstein}},
  \bibinfo{journal}{Physical review letters} \textbf{\bibinfo{volume}{116}},
  \bibinfo{pages}{240405} (\bibinfo{year}{2016}).

\bibitem[{\citenamefont{Zhu et~al.}(2017)\citenamefont{Zhu, Ma, Cao, Fei, and
  Vedral}}]{zhu_2017_operational}
\bibinfo{author}{\bibfnamefont{H.}~\bibnamefont{Zhu}},
  \bibinfo{author}{\bibfnamefont{Z.}~\bibnamefont{Ma}},
  \bibinfo{author}{\bibfnamefont{Z.}~\bibnamefont{Cao}},
  \bibinfo{author}{\bibfnamefont{S.-M.} \bibnamefont{Fei}}, \bibnamefont{and}
  \bibinfo{author}{\bibfnamefont{V.}~\bibnamefont{Vedral}},
  \bibinfo{journal}{Physical Review A} \textbf{\bibinfo{volume}{96}},
  \bibinfo{pages}{032316} (\bibinfo{year}{2017}).

\bibitem[{\citenamefont{Chitambar et~al.}(2016)\citenamefont{Chitambar,
  Streltsov, Rana, Bera, Adesso, and Lewenstein}}]{chitambar_2016_assisted}
\bibinfo{author}{\bibfnamefont{E.}~\bibnamefont{Chitambar}},
  \bibinfo{author}{\bibfnamefont{A.}~\bibnamefont{Streltsov}},
  \bibinfo{author}{\bibfnamefont{S.}~\bibnamefont{Rana}},
  \bibinfo{author}{\bibfnamefont{M.}~\bibnamefont{Bera}},
  \bibinfo{author}{\bibfnamefont{G.}~\bibnamefont{Adesso}}, \bibnamefont{and}
  \bibinfo{author}{\bibfnamefont{M.}~\bibnamefont{Lewenstein}},
  \bibinfo{journal}{Physical review letters} \textbf{\bibinfo{volume}{116}},
  \bibinfo{pages}{070402} (\bibinfo{year}{2016}).

\bibitem[{\citenamefont{Buscemi and Datta}(2013)}]{buscemi_2013_general}
\bibinfo{author}{\bibfnamefont{F.}~\bibnamefont{Buscemi}} \bibnamefont{and}
  \bibinfo{author}{\bibfnamefont{N.}~\bibnamefont{Datta}},
  \bibinfo{journal}{IEEE Transactions on Information Theory}
  \textbf{\bibinfo{volume}{59}}, \bibinfo{pages}{1940} (\bibinfo{year}{2013}).

\bibitem[{\citenamefont{Regula et~al.}(2017)\citenamefont{Regula, Fang, Wang,
  and Adesso}}]{regula_2017_one}
\bibinfo{author}{\bibfnamefont{B.}~\bibnamefont{Regula}},
  \bibinfo{author}{\bibfnamefont{K.}~\bibnamefont{Fang}},
  \bibinfo{author}{\bibfnamefont{X.}~\bibnamefont{Wang}}, \bibnamefont{and}
  \bibinfo{author}{\bibfnamefont{G.}~\bibnamefont{Adesso}},
  \bibinfo{journal}{arXiv preprint arXiv:1711.10512}  (\bibinfo{year}{2017}).

\bibitem[{\citenamefont{Zhao et~al.}(2018)\citenamefont{Zhao, Liu, Yuan,
  Chitambar, and Ma}}]{zhao_2018_one}
\bibinfo{author}{\bibfnamefont{Q.}~\bibnamefont{Zhao}},
  \bibinfo{author}{\bibfnamefont{Y.}~\bibnamefont{Liu}},
  \bibinfo{author}{\bibfnamefont{X.}~\bibnamefont{Yuan}},
  \bibinfo{author}{\bibfnamefont{E.}~\bibnamefont{Chitambar}},
  \bibnamefont{and} \bibinfo{author}{\bibfnamefont{X.}~\bibnamefont{Ma}},
  \bibinfo{journal}{Physical Review Letters} \textbf{\bibinfo{volume}{120}},
  \bibinfo{pages}{070403} (\bibinfo{year}{2018}).

\bibitem[{\citenamefont{Zhao and Winter}(Manuscript in
  preperation)}]{Zhao_winter_2018}
\bibinfo{author}{\bibfnamefont{Q.}~\bibnamefont{Zhao}} \bibnamefont{and}
  \bibinfo{author}{\bibfnamefont{A.}~\bibnamefont{Winter}}
  (\bibinfo{year}{Manuscript in preperation}).

\bibitem[{\citenamefont{Bowen and Datta}(2006)}]{bowen_2006_beyond}
\bibinfo{author}{\bibfnamefont{G.}~\bibnamefont{Bowen}} \bibnamefont{and}
  \bibinfo{author}{\bibfnamefont{N.}~\bibnamefont{Datta}}, in
  \emph{\bibinfo{booktitle}{Information Theory, 2006 IEEE International
  Symposium on}} (\bibinfo{organization}{IEEE}, \bibinfo{year}{2006}), pp.
  \bibinfo{pages}{451--455}.

\bibitem[{\citenamefont{Ogawa and Nagaoka}(2002)}]{ogawa_2002_new}
\bibinfo{author}{\bibfnamefont{T.}~\bibnamefont{Ogawa}} \bibnamefont{and}
  \bibinfo{author}{\bibfnamefont{H.}~\bibnamefont{Nagaoka}}, in
  \emph{\bibinfo{booktitle}{Information Theory, 2002. Proceedings. 2002 IEEE
  International Symposium on}} (\bibinfo{organization}{IEEE},
  \bibinfo{year}{2002}), p.~\bibinfo{pages}{73}.

\bibitem[{\citenamefont{Winter}(1999)}]{winter_1999_coding}
\bibinfo{author}{\bibfnamefont{A.}~\bibnamefont{Winter}},
  \bibinfo{journal}{IEEE Transactions on Information Theory}
  \textbf{\bibinfo{volume}{45}}, \bibinfo{pages}{2481} (\bibinfo{year}{1999}).

\bibitem[{\citenamefont{Sion}(1958)}]{Sion-1958a}
\bibinfo{author}{\bibfnamefont{M.}~\bibnamefont{Sion}},
  \bibinfo{journal}{Pacific Journal of Mathematics}
  \textbf{\bibinfo{volume}{8}}, \bibinfo{pages}{171–} (\bibinfo{year}{1958}).

\bibitem[{\citenamefont{Alicki and Fannes}(2004)}]{Alicki_2004_continuity}
\bibinfo{author}{\bibfnamefont{R.}~\bibnamefont{Alicki}} \bibnamefont{and}
  \bibinfo{author}{\bibfnamefont{M.}~\bibnamefont{Fannes}},
  \bibinfo{journal}{Journal of Physics A: Mathematical and General}
  \textbf{\bibinfo{volume}{37}}, \bibinfo{pages}{L55} (\bibinfo{year}{2004}).

\bibitem[{\citenamefont{Datta}(2009)}]{datta_2009_min}
\bibinfo{author}{\bibfnamefont{N.}~\bibnamefont{Datta}}, \bibinfo{journal}{IEEE
  Transactions on Information Theory} \textbf{\bibinfo{volume}{55}},
  \bibinfo{pages}{2816} (\bibinfo{year}{2009}).

\end{thebibliography}

\appendix 

\section{Proof of equation (\ref{qc_c_r_eqn})}\label{eqn_proof_appendix}

\begin{equation} \label{eqn_to_prove}
\begin{split}
C_{r}\left( \sum\limits_i p_i \phi_i \otimes \pi_i \right) &= S\left( \Delta\left( \sum\limits_i p_i \phi_i \otimes \pi_i  \right) \right) - S\left( \sum\limits_i p_i \phi_i \otimes \pi_i \right) \\
&= S \left( \sum\limits_i p_i \Delta (\phi_i) \otimes \pi_i  \right) - S \left( \sum\limits_i p_i \phi_i \otimes \pi_i  \right)
\end{split}
\end{equation}
For a general quantum-classical state, $\sigma = \sum\limits_i q_i \sigma_i \otimes \pi_i$,
\begin{equation}
\begin{split}
S(\sigma) &= -\Tr\sigma\ln\sigma \\ 
&= -\Tr \left( \left( \sum\limits_i p_i \sigma_i \otimes \pi_i \right) \ln \left( \sum\limits_j p_j \sigma_j \otimes \pi_j  \right) \right)  \\
&= -\Tr \left( \left( \sum\limits_{i, k} p_i \lambda_k^i \ketbra{\lambda_k^i}{\lambda_k^i} \otimes \pi_i \right) \ln \left( \sum\limits_{j,l} p_j \lambda_l^j \ketbra{\lambda_l^j}{\lambda_l^j} \otimes \pi_j  \right) \right) \\
&= -\Tr \left( \sum\limits_{i, k}  p_i \lambda_k^i \ln( p_i \lambda_k^i) \ketbra{\lambda_k^i}{\lambda_k^i} \otimes \pi_i  \right) \\
&=  -\sum\limits_{i, k}  p_i \lambda_k^i \ln( p_i \lambda_k^i) \\
&=  -\sum\limits_{i, k}  p_i \lambda_k^i \ln( p_i) -  \sum\limits_{i, k}  p_i \lambda_k^i \ln(  \lambda_k^i) \\
&= -\sum\limits_i p_i \ln p_i + \sum_i p_i S(\sigma_i)
\end{split}
\end{equation}

Applying the above result to equation \eqref{eqn_to_prove} we get,
\begin{equation} 
\begin{split}
C_{r}\left( \sum\limits_i p_i \phi_i \otimes \pi_i \right) &= \sum\limits_{i}p_i S(\Delta(\phi_i)) \\
&= \sum\limits_{i}p_i C_r(\phi_i)
\end{split}
\end{equation}

\section{$\mathcal{M}$ satisfies generalized Stein's lemma} \label{stein_appendix}
For the generalized Stein's lemma to hold for a family of sets $\mathcal{M}$ the following conditions need to be met \cite{brandao_2010_generalization}
\begin{enumerate} 
\item Each $\mathcal{M}_n$ must be closed and convex.
\item Each $\mathcal{M}_n$ contains $\sigma^{\otimes n}$ for a full rank state $\sigma \in \mathcal{D}(\mathcal{H})$.
\item If $\rho \in \mathcal{M}_{n + 1}$, then $\Tr_k(\rho) \in \mathcal{M}_n$, for every $k \in \lbrace 1,...,n+1 \rbrace$.
\item If $\rho \in \mathcal{M}_n$ and $\nu \in \mathcal{M}_m$, then $\rho \otimes \nu \in \mathcal{M}_{n + m}$.
\item If $\rho \in \mathcal{M}_n$ then $P_{\pi}\rho P_{\pi} \in \mathcal{M}_n $ for every $\pi \in S_n$, where $P_{\pi}$ is the representation of a permutation $\pi$ in $\mathcal{H}^{\otimes n}$ and $S_n$ is symmetric group of order $n$.
\end{enumerate}

The set of incoherent states in $\mathcal{H}^{\otimes n}$ will be convex and closed satisfying the first condition.  $\delta^{\otimes n} \in \mathcal{I}_n$ satisfying condition 2. $\Tr_k(\delta_{n + 1}) \in \mathcal{I}_n$ where $\delta_{n + 1} \in \mathcal{I}_{n + 1 }$ for any $k \in \lbrace 1,...,n+1 \rbrace$ satisfying condition 3. $\delta_n \otimes \nu_m \in \mathcal{I}_{m +n}$ when $\delta_n \in \mathcal{I}_n$ and $\nu_m \in \mathcal{I}_m$ hence condition 4. is satisfied. Finally the permutation operation is just a relabelling of the incoherent basis hence the set of incoherent states will be closed under such a permutation and condition 5. is satisfied.

  % input acknowledgement

\end{document}